\documentclass[twocolumn,showpacs,amsmath,amssymb,aps,pra,floatfix]{revtex4}
\usepackage{bm}
\usepackage{amsfonts}
\usepackage{amsmath}
\usepackage{graphicx}
\usepackage{epstopdf}

\begin{document}
\title{Transport of the light polarization in the weak gravitational wave background}
\author{Piotr Bargiela}
\email{piotr.bargiela@gmail.pl}
\affiliation{Center for Theoretical Physics, Polish Academy of Sciences, Al. Lotnik\'ow 32/46, 02-668 Warsaw, Poland}

\begin{abstract}
The influence of the weak gravitational wave on the light polarization is considered. Oscillations in the direction of the polarization vector is found.
\end{abstract}
\pacs{04.30.-w, 42.25.Ja} \maketitle

\section{Introduction}
There is a well understood and developed method to measure the parameters of passing gravitational radiation by interferometers. It is based on using the light as a medium to distinguish the fluctuations in the distance between two far points. The method introduced here treats the light not only as a medium but as a physical indicator of the appearance of the gravitational wave.

Firstly, general polarization transport equation is derived. Riemann-Silberstein formalism is used. Secondly, weak plane gravitational wave metric is diagonalized. Thirdly, geodesic equation for massless particle is solved in new orthogonal coordinates. Finally, polarization transport equations are solved along obtained geodesic. Oscillations in the direction of the light polarization vector can be noticed.

\section{Geometrical optics}
\label{GO}
In order to consider light polarization transport in curved spacetime, it is needed to begin with the Maxwell equations. The description of electromagnetic waves is significantly easier when using the Riemann-Silberstein formalism. Let us define the relativistic antisymmetric complex tensor \cite{Birula}
\begin{align}\label{RStensor}
F^{\mu\nu}=\sqrt{-g}\left(g^{\mu\lambda}g^{\nu\rho}+
\frac{i}{2\sqrt{-g}}\varepsilon^{\mu\nu\lambda\rho}\right)f_{\lambda\rho} \ ,
\end{align}
where $f_{\mu\nu}=\partial_\mu A_\nu-\partial_\nu A_\mu$ is real electromagnetic field tensor.
The relativistic tensor $F^{\mu\nu}$ has three independent components and is self-dual
\begin{align}\label{dual}
\frac{i}{2\sqrt{-g}}\varepsilon^{\mu\nu}_{\ \ \lambda\rho}F^{\lambda\rho}=F^{\mu\nu} \ .
\end{align}
The Maxwell equations in terms of the relativistic tensor $F^{\mu\nu}$ yield
\begin{align}\label{Maxwell}
\partial_\mu F^{\mu\nu}=0 \ .
\end{align}

With regards to the light propagation, for large value of the light angular frequency $\Omega$, it is convenient to use eikonal form of the relativistic tensor $F^{\mu\nu}$
\begin{align}\label{Eikonal}
F^{\mu\nu}=e^{i\Omega S}\mathcal{F}^{\mu\nu} \ .
\end{align}
This is a consequence of the WKB approximation \cite{MTW}, which enables the passage from electromagnetic field dynamics to the propagation of rays with polarization. The wave vector $k^\mu$ is defined as normal to the surface of constant phase $\psi=\Omega S$. From the Maxwell equations one can find
\begin{align}\label{kMaxwell}
k_\mu \mathcal{F}^{\mu\nu}=0 \ .
\end{align}
Due to the source-free wave equation,
the wave vector $k_\mu$ is null
\begin{align}\label{knull}
k_\mu k^\mu=0
\end{align}
and immediately $S$ obeys the eikonal equation
\begin{align}\label{eikonalEq}
\frac{\partial S}{\partial x^\mu}\frac{\partial S}{\partial x_\mu}=0 \ .
\end{align}
Taking covariant derivative of (\ref{knull}) and choosing photon trajectory $k^\mu=\frac{dx^\mu}{d\lambda}$ leads to the statement that the wave vector moves parallel along the geodesic
\begin{align}\label{kParallel}
\frac{dk^\mu}{d\lambda}+\Gamma^\mu_{\nu\sigma}k^\nu k^\sigma=0 \ .
\end{align}

Seeking transport of the polarization, let us represent tensor amplitude in the form
\begin{align}\label{tensAmp}
\mathcal{F}^{\mu\nu}=a\epsilon^{\mu\nu} \ ,
\end{align}
where $a=(\mathcal{F}^{\mu\nu}\mathcal{F}^*_{\mu\nu})^{1/2}$ is the scalar amplitude and $\epsilon^{\mu\nu}$ is the polarization tensor. The tensor amplitude $\mathcal{F}^{\mu\nu}$ propagates according to
\begin{align}\label{Fprop}
k^\delta_{\ ;\delta}\mathcal{F}_{\mu\nu}+2k^\delta\mathcal{F}_{\mu\nu;\delta}=0 \ ,
\end{align}
the scalar amplitude $a$ obeys
\begin{align}\label{aprop}
k^\delta_{\ ;\delta}a+2k^\delta a_{;\delta}=0
\end{align}
and the polarization tensor $\epsilon^{\mu\nu}$ moves parallel along the geodesic
\begin{align}\label{eprop}
\frac{d\epsilon^{\mu\nu}}{d\lambda}+\Gamma^\mu_{\delta\sigma}k^\delta\epsilon^{\sigma\nu}+
\Gamma^\nu_{\delta\sigma}k^\delta\epsilon^{\mu\sigma}=0 \ .
\end{align}
This is a general result, which allows us to analyse our main problem.

\section{Gravitational wave}
\label{GW}
Let us consider linear correction
\begin{align}\label{corrention}
h_{\mu\nu}=\begin{pmatrix}
  0 & 0 & 0 & 0 \\
  0 & h_+ \cos{\varphi} & h_\times \cos{\varphi} & 0 \\
  0 & h_\times \cos{\varphi} & -h_+ \cos{\varphi} & 0 \\
  0 & 0 & 0 & 0
 \end{pmatrix}
\end{align}
to the flat Minkowski metric tensor $\eta_{\mu\nu}$ with signature $(+,-,-,-)$. We denoted $\varphi=kz-\omega t$, where $k$ is the wavenumber and $\omega=ck$ is the angular frequency of the gravitational wave. By the subtraction $g_{\mu\nu}=\eta_{\mu\nu}-h_{\mu\nu}$ we get the line element
\begin{align}\label{metricWave}
ds^2=c^2 dt^2-(1+h_+ \cos{\varphi})dx^2-2h_\times \cos{\varphi}dxdy- \notag \\
(1-h_+ \cos{\varphi})dy^2-dz^2 \ .
\end{align}

Since this metric tensor is non-diagonal, it would be more convenient to have orthogonal coordinate system. As it turns out, it is possible by choosing
$\sigma_1=\sigma_4=1$,
$\sigma_2=\frac{1}{h_\times\sqrt{2}}\sqrt{1-\frac{h_+}{H}}$,
$\sigma_3=\frac{1}{\sqrt{2}h_\times}\sqrt{1+\frac{h_+}{H}}$
in eigenvectors
\begin{align}\label{eigenvectors}
\begin{cases}
\chi_1=(1,0,0,0)\sigma_1\\
\chi_2=(0,h_+ + H,h_\times,0)\sigma_2\\
\chi_3=(0,h_+ - H,h_\times,0)\sigma_3\\
\chi_4=(0,0,0,1)\sigma_4
\end{cases} \ ,
\end{align}
where $H=\sqrt{h_+^2+h_\times^2}$. Corresponding transformation of coordinates
\begin{align}\label{newCoord}
\begin{cases}
t=t'\\
x=(h_+ + H)\sigma_2 x'+(h_+ - H)\sigma_3 y'\\
y=h_\times\sigma_2 x'+h_\times\sigma_3 y'\\
z=z'
\end{cases} \ .
\end{align}
Diagonal metric tensor has a form
\begin{align}\label{metricDiag}
\tilde{g}_{\mu\nu}=\begin{pmatrix}
  1 & 0 & 0 & 0 \\
  0 & -(1+H \cos{\varphi}) & 0 & 0 \\
  0 & 0 & -(1-H \cos{\varphi}) & 0 \\
  0 & 0 & 0 & -1
 \end{pmatrix} \ .
\end{align}
To simplify later calculations we will avoid prime symbol (') in the higher index of the new coordinates.

All following calculations are appropriate only in the first order $\mathcal{O}(H^1)$ with regards to the construction of the metric linear correction (\ref{corrention}).

\section{Photon trajectory}
\label{PT}
Every particle, even massless, moves along a geodesic line in the curved spacetime \cite{Weinberg}. Geodesic equation yields
\begin{align}\label{geodesicEq}
\frac{d^2x^\mu}{d\lambda^2}+\Gamma^\mu_{\nu\sigma}\frac{dx^\nu}{d\lambda}\frac{dx^\sigma}{d\lambda}=0 \ ,
\end{align}
where
$\Gamma^\mu_{\nu\sigma}=\frac{1}{2}\tilde{g}^{\mu\rho}(\tilde{g}_{\rho\nu,\sigma}+
\tilde{g}_{\rho\sigma,\nu}-\tilde{g}_{\nu\sigma,\rho})$
are the Christoffel symbols of the second kind and $\lambda$ is the affine parameter. Evaluated geodesic equations in the metric (\ref{metricDiag}) are
\begin{align}\label{geodesicEqEval}
\begin{cases}
c^2\ddot{t}+\frac{1}{2}H\omega(\dot{x}^2-\dot{y}^2)\sin{\varphi}=0 \\
(1+H\cos{\varphi})\ddot{x}-H\dot{x}\dot{\varphi}\sin{\varphi}=0 \\
(1-H\cos{\varphi})\ddot{x}+H\dot{y}\dot{\varphi}\sin{\varphi}=0 \\
\ddot{z}-\frac{1}{2}Hk(\dot{y}^2-\dot{x}^2)\sin{\varphi}=0
\end{cases} \ ,
\end{align}
where $\dot{x^\mu}=\frac{dx^\mu}{d\lambda}$. These equations can be integrated directly
\begin{align}\label{geodesicEqSimp}
\begin{cases}
(1+H\cos{\varphi})\dot{x}=const=p_1 \\
(1-H\cos{\varphi})\dot{y}=const=p_2 \\
\dot{z}+\frac{Hk}{2p_0}(p_2^2-p_1^2)\cos{\varphi}=const=p_3 \\
c\dot{t}+\frac{Hk}{2p_0}(p_2^2-p_1^2)\cos{\varphi}=const=p_4
\end{cases}
\end{align}
with $k\dot{z}-\omega\dot{t}=const=p_0$. First pair of equations is closely related to the fact that the Lagrangian $\mathcal{L}=\frac{1}{2}\tilde{g}_{\mu\nu}d\dot{x}^\mu d\dot{x}^\nu$ is independent of the $x$ and $y$ variable. Second pair of equations provides that there are only four independent constants of motion $p_4=p_3-\frac{p_0}{k}$. In consequence one can get
\begin{align}\label{geodetic}
\begin{cases}
x(\lambda)=\frac{p_1}{p_0}[\lambda+H(\sin{\varphi_0}-\sin{\varphi(\lambda)})]+x_0 \\
y(\lambda)=\frac{p_2}{p_0}[\lambda-H(\sin{\varphi_0}-\sin{\varphi(\lambda)})]+y_0 \\
z(\lambda)=p_3\lambda+H\frac{k(p_1^2-p_2^2)}{p_0^2}\cos{\frac{p_0\lambda}{2}}
\sin(\frac{p_0\lambda}{2}+\varphi_0)+z_0 \\
t(\lambda)=\frac{p_4}{c^2}\lambda+H\frac{k(p_1^2-p_2^2)}{c^2p_0^2}\cos{\frac{p_0\lambda}{2}}
\sin(\frac{p_0\lambda}{2}+\varphi_0)+t_0
\end{cases} \ ,
\end{align}
where $\varphi(\lambda)=p_0\lambda+\varphi_0$.

It can be observed, that every function has a linear in $\lambda$ component and a periodic correction proportional to $H$. The most interesting relation, in the view of calculating the classical position vector $\textbf{r}(t)$, $t(\lambda)$ cannot be simply inverted. However, fully analytic form of the velocity can be obtained using the chain rule $\textbf{v}(t)=\frac{d\textbf{r}(\lambda)}{d\lambda}\frac{d\lambda(t)}{dt}$ and the inverse function derivative $\frac{d\lambda(t)}{dt}=\frac{1}{\frac{dt(\lambda)}{d\lambda}}$. Of course, the trajectory of the particle can be parameterized by any parameter, $\lambda$ in particular.

Well known property, that trajectory of the particle moving only in the propagation direction of the gravitational wave is not disturbed, is clear in (\ref{geodetic}). In this case $p_1=p_2=0$, then $z(\lambda)$ and $t(\lambda)$ are linear in $\lambda$.

Line element
\begin{align}\label{NullGeodesic}
ds^2=\left(p_2^2-p_1^2-\frac{2p_0p_3}{k}-\frac{p_0^2}{k^2}\right)d\lambda^2
\end{align}
is larger than zero for massive particles. For photons it equals zero, therefore the null geodesic condition is
\begin{align}\label{NullGeodesicCond}
k^2(p_2^2-p_1^2)-2kp_0p_3-p_0^2=0 \ .
\end{align}

\section{Polarization transport}
\label{PT}
Dealing with solving parallel transport equations (\ref{eprop}), it is useful to reduce polarization tensor
$\epsilon^{\mu\nu}$. Let us introduce polarization vector $\epsilon^\mu$
\begin{align}\label{polarVect}
ae^{i\Omega S}\epsilon^\mu=n_\nu F^{\mu\nu}
\end{align}
with arbitrary timelike vector $n_\nu$. The polarization vector is perpendicular to the wave vector $k_\mu\epsilon^\mu=0$ and moves parallel along the geodesic like its tensor counterpart $\epsilon^{\mu\nu}$. We choose $n_\nu$ in the simplest way $n_\nu=(1,0,0,0)$ to obtain
\begin{align}\label{polVectN}
ae^{i\Omega S}\epsilon^\mu=F^{\mu0}=(0,\textbf{F}) \ ,
\end{align}
where the Riemann-Silberstein vector $\textbf{F}$ is a complex combination of electromagnetic field vectors
\begin{align}\label{RSvector}
\textbf{F}=\frac{\textbf{D}}{\sqrt{2\epsilon}}+i\frac{\textbf{B}}{\sqrt{2\mu}}
\end{align}
with dielectric constant $\epsilon$ and magnetic permeability $\mu$.

The parallel transport of the vector $\epsilon^\mu$ along the geodesic line yields
\begin{align}\label{transportEq}
\frac{d\epsilon^\mu}{d\lambda}+\Gamma^\mu_{\nu\sigma}\frac{dx^\nu}{d\lambda}\epsilon^\sigma=0 \ .
\end{align}
Using (\ref{geodetic}), the expanded form of the above equations is
\begin{align}\label{transportEqEval}
\begin{cases}
c^2\dot{\epsilon}^t+\frac{1}{2}H\omega(\epsilon^x\dot{x}-\epsilon^y\dot{y})\sin{\varphi}=0 \\
(1+H\cos{\varphi})\dot{\epsilon}^x \ - \\
\ \ \ \ \frac{1}{2}H[\epsilon^x\dot{\varphi}+\dot{x}(k\epsilon^z-\omega \epsilon^t)]\sin{\varphi}=0 \\
(1-H\cos{\varphi})\dot{\epsilon}^y \ + \\
\ \ \ \ \frac{1}{2}H[\epsilon^y\dot{\varphi}+\dot{y}(k\epsilon^z-\omega \epsilon^t)]\sin{\varphi}=0 \\
\dot{\epsilon}^z-\frac{1}{2}Hk(\epsilon^y\dot{y}-\epsilon^x\dot{x})\sin{\varphi}=0
\end{cases} \ .
\end{align}
The general solutions are
\begin{align}\label{genSol}
\begin{cases}
\epsilon^x(\lambda)=\frac{1}{2}H\left(\frac{p_1q_0}{p_0^2}+\epsilon^x_0\right)
(\cos{\varphi_0}-\cos{\varphi(\lambda)})+\epsilon^x_0 \\
\epsilon^y(\lambda)=-\frac{1}{2}H\left(\frac{p_2q_0}{p_0^2}+\epsilon^y_0\right)
(\cos{\varphi_0}-\cos{\varphi(\lambda)})+\epsilon^y_0 \\
\epsilon^z(\lambda)=\frac{k}{p_0^2}H(\epsilon^y_0p_2-\epsilon^x_0p_1)\sin{\frac{p_0\lambda}{2}}
\sin\left(\frac{p_0\lambda}{2}+\varphi_0\right)+\epsilon^z_0 \\
\epsilon^t(\lambda)=\frac{k}{cp_0^2}H(\epsilon^y_0p_2-\epsilon^x_0p_1)\sin{\frac{p_0\lambda}{2}}
\sin\left(\frac{p_0\lambda}{2}+\varphi_0\right)+\epsilon^t_0
\end{cases} \ ,
\end{align}
where $q_0=k\epsilon^z-\omega\epsilon^t=const=k\epsilon^z_0-\omega\epsilon^t_0$.

According to the definition of $\epsilon^\mu$, $\epsilon^0=\epsilon^t=0$, then initial conditions $\epsilon^\mu_0$ must obey $\epsilon^x_0p_1=\epsilon^y_0p_2$ and $\epsilon^t_0=0$. Therefore
$\epsilon^z=\epsilon^z_0$. Finally, solutions of the polarization vector parallel transport equations are
\begin{align}\label{polarSol}
\begin{cases}
\epsilon^x(\lambda)=\frac{1}{2}H\left(k\frac{p_1}{p_0^2}\epsilon^z_0+\epsilon^x_0\right)
(\cos{\varphi_0}-\cos{\varphi(\lambda)})+\epsilon^x_0 \\
\epsilon^y(\lambda)=-\frac{1}{2}H\left(k\frac{p_2}{p_0^2}\epsilon^z_0+\epsilon^y_0\right)
(\cos{\varphi_0}-\cos{\varphi(\lambda)})+\epsilon^y_0 \\
\epsilon^z(\lambda)=\epsilon^z_0 \\
\epsilon^t(\lambda)=0
\end{cases} \ .
\end{align}

In order to construct the unit polarization vector
\begin{align}\label{length}
-\epsilon^\mu\epsilon^*_\mu=|\epsilon^x_0|^2+|\epsilon^y_0|^2+|\epsilon^z_0|^2+
H\left(|\epsilon^x_0|^2-|\epsilon^y_0|^2\right)\cos{\varphi_0}=1
\end{align}
one can choose
\begin{align}\label{ez0}
\epsilon^z_0=\left(\sqrt{1-|\epsilon^x_0|^2-|\epsilon^y_0|^2}-
H\frac{|\epsilon^x_0|^2-|\epsilon^y_0|^2}{2\sqrt{1-|\epsilon^x_0|^2-|\epsilon^y_0|^2}}\cos{\varphi_0}\right)e^{i\phi}
\end{align}
with arbitrary real $\phi$.

\section{Summary}
Because of the existence of the periodic factors in (\ref{geodetic}) and (\ref{polarSol}), there is a tiny oscillation in the direction of the light polarization vector $\epsilon^\mu$, which is proportional to the strength $H$ of the gravitational wave. It can be used as a new way to measure the parameters of the passing gravitational radiation without building large-scale interferometers.

\section*{Acknowledgments}
I would like to thank to Professor Iwo Bialynicki-Birula for lots of insightful discussions and suggestions. The work was supported by Polish National Science Center Grant No. 2012/07/B/ST1/03347.

\end{document}